\documentclass[useAMS,usenatbib]{mnras}

\usepackage{amsmath}
\usepackage{amssymb}
\usepackage{graphicx}
\usepackage{times}

\title[Impact of filaments on galaxy formation]
  {Impact of filaments on galaxy formation in their residing dark matter haloes}

\author[Liao \& Gao]
{Shihong Liao$^{1}$ \thanks{Email: shliao@nao.cas.cn} and Liang Gao$^{1,2}$ 
\\
$^1$Key Laboratory for Computational Astrophysics, National
Astronomical Observatories, Chinese Academy of Sciences, Beijing,
100012, China\\
$^2$Institute of Computational Cosmology, Department of Physics,
University of Durham, Science Laboratories, South Road, Durham DH1
3LE, UK\\
}
\begin{document}



\maketitle

\label{firstpage}

\begin{abstract}
We make use of a high-resolution zoom-in hydrodynamical simulation to investigate the impact of filaments on galaxy formation in their residing dark matter haloes. A method based on the density field and the Hoshen-Kopelman algorithm is developed to identify filaments. We show that cold and dense gas preprocessed by dark matter filaments can be further accreted into residing individual low-mass haloes in directions along the filaments. Consequently, comparing with field haloes, gas accretion is very anisotropic for filament haloes. About 30 percent of the accreted gas of a residing filament halo was preprocessed by filaments, leading to two different thermal histories for the gas in filament haloes. Filament haloes have higher baryon and stellar fractions when comparing with their field counterparts. Without including stellar feedback, our results suggest that filaments assist gas cooling and enhance star formation in their residing dark matter haloes at high redshifts (i.e. $z=4.0$ and $2.5$).
\end{abstract}


\section{Introduction}\label{sec_intro}
In the classic galaxy formation framework \citep[e.g.,][]{white1978, white1991}, to form luminous galaxies, primordial gas should fall into the potential wells created by dark matter haloes, be shock heated and become ionized, cool radiatively, and form stars. Comparing with dark matter haloes, massive filaments have the same function to provide potential wells to trap gas. As demonstrated in \citet[][]{gao2007} and \citet[][]{gao2015}, gas accretion into filaments can be very similar to what happens in dark matter haloes in warm dark matter (WDM) models, namely that the gas is shock heated at the boundary of filaments, cools radiatively and condenses into the filament centres to form stars. 

While the above two studies mainly focus on gas physics in filaments in WDM models, the same physics should of course apply to cold dark matter (CDM) models because the large scale structures of filaments are quite similar in both models. As shown in a hydrodynamical simulation of \citet[][]{gao2015} and many other studies \citep[e.g.][]{keres2005, dekel2006}, there indeed exists cold and dense gas flows, preprocessed by dark matter filaments, in the centres of CDM filaments at high redshifts. The phenomenon that cold and dense gas flows fuel massive central galaxies is termed ``cold accretion'', and it has been extensively studied in many literatures \citep[e.g.][]{keres2005, dekel2006, ocvirk2008, brooks2009, dekel2009, keres2009, voort2011, danovich2012}. However, how the filament environment affects the formation of residing galaxies in filament itself has been poorly explored, which is the main focus of this study.

Recently, there are more and more evidence unveiling the importance of filaments on galaxy properties in observations. For instance, when compared with other environments (e.g. clusters and fields), galaxies in filaments have enhanced star formation activities \citep[e.g.][]{porter2007, fadda2008, porter2008, biviano2011, coppin2012, darvish2014}, tend to be more metal enriched \citep[][]{darvish2015}, and have more satellites \citep[][]{guo2015}, etc. When studying galaxies' properties as a function of their distances to the nearest filament, \citet[][]{alpaslan2016} shown that galaxies in the GAMA survey, with stellar masses $M_\star > 10^{9}$ $h^{-2}M_\odot$ and redshifts $z<0.09$, tend to have higher stellar masses and lower star formation rates when they reside closer to the cylindrical centre of a filament. Similar conclusions were seen by galaxies with higher redshifts $(z \lesssim 1.0)$ in the GAMA survey \citep[][]{kraljic2018}, the VIPERS survey \citep[][]{malavasi2017}, the COSMOS2015 galaxy catalogue \citep[][]{darvish2017, laigle2017} and the SDSS survey \citep[][]{chen2017, kuutma2017, poudel2017}.

In this study, we use a high-resolution cosmological hydrodynamical simulation to study the gas accretion and star formation of low-mass haloes residing in filaments formed at high redshifts within the Lambda cold dark matter ($\Lambda$CDM) model. The paper is structured as follows. We overview the details of the simulation, and present our filament identification methods in Section \ref{sec_met}. The detailed gas accretion process in residing low-mass haloes of  filaments, and the impacts of filaments on galaxy formaition are discussed in Section \ref{sec_res}. Section \ref{sec_con} summarizes our results.

\section{Methods} \label{sec_met}
\subsection{Numerical simulations} \label{subsec_sim}

\begin{figure*} 
\centering\includegraphics[width=370pt]{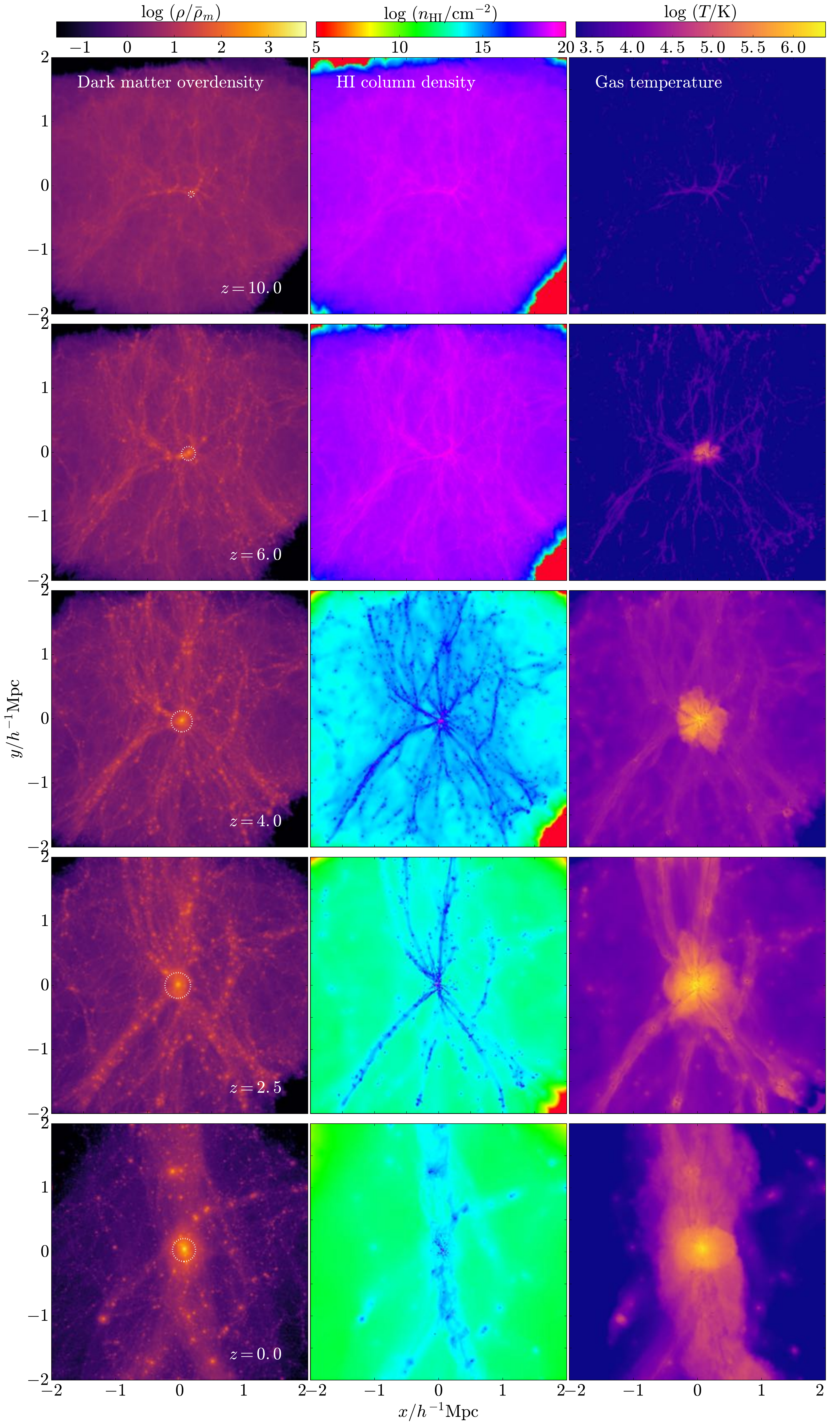} 
\caption{Visualization of the dark matter overdensity (left), HI column density (middle) and gas temperature (right) in our simulation at $z=10.0, 6.0, 4.0, 2.5,$ and $0.0$ (from top to bottom). At each redshift, we select a comoving cubic region centering on the center-of-mass of the zoom-in region and project it on the $xy$ plane. The edge length of the cube is $4$ $h^{-1}\mathrm{Mpc}$. The dotted circles in the first column mark the position and virial radius of the progenitor of the Aq-A halo. The color bars are shown on the top of each column. We can see several massive filaments form at $z=4.0$ and penetrate into the Aq-A halo, and become more massive at $z=2.5$. We will focus on the gas accretion and halo formation inside these filaments in later discussions.}\label{fig_sim_visua}
\end{figure*}

The simulation used in this study is a high-resolution hydrodynamical re-simulation of an individual galactic sized halo, Aq-A, from the Aquarius project \citep[][]{springel2008}. This simulation had been used to study the star-forming filaments in \citet[][]{gao2015}. The cosmological parameters adopted in the simulation are $\Omega_m=0.25$, $\Omega_b=0.045$, $\Omega_\Lambda=0.75$, $\sigma_8=0.9$ and $h=0.73$.  The mass resolution of dark matter particles in the high-resolution region of the simulation is $m_\mathrm{DM}=2.35\times 10^5$ $h^{-1}M_\odot$, and it is $m_\mathrm{gas}=5.16\times 10^4$ $h^{-1}M_\odot$ for each gas particle. The force resolution is $\epsilon=0.5$ $h^{-1}\mathrm{kpc}$ in comoving units.

The simulation includes radiative cooling and photo-heating of a primordial gas in the presence of a UV/X-ray background from galaxies and quasars \citep[][]{haardt1996}. The simulation assumes that the HI ionization occurs at $z=6$. To model star formation, the simulation converts gas into stars with a physical hydrogen density threshold of $n_\mathrm{H}=0.1$ $\mathrm{cm}^{-3}$ and an overdensity threshold of $\rho_\mathrm{gas}/\bar{\rho}_\mathrm{gas}=2000$. Note that, as a first step, this study focuses on the investigation of gas accretion and cooling in filaments and their residing low-mass dark matter haloes, and feedbacks from supernovae and active galactic nuclei are not included in the simulation. The simulation has been evolved from $z=127$ to the present day using \textsc{gadget}-3 code \cite[][]{springel2005}. 

We adopt the Amiga Halo Finder \citep[\textsc{ahf},][]{knollmann2009} to identify haloes in the high-resolution region of our simulation with a virial parameter of $\Delta_\mathrm{vir}=200$  measured with respect to the critical density. Note that in the remaining of this article, we simply call an \textsc{ahf} identified structure as a ``halo" instead of a ``galaxy", although it may contain gas and stars. We have identified all individual haloes with more than $10$ particles, but only adopt those with at least $100$ particles to carry out our analysis. The mass range of our halo catalogue is approximately $[10^{7.3}, 10^{11}]h^{-1}M_\odot$. To avoid possible influences from the central Aq-A halo and the boundary of the zoom-in region, we only use haloes whose comoving distances from the centre of the Aq-A halo, $r$, satisfy $2R_{200,\mathrm{AqA}}(z)<r<2$ $h^{-1}\mathrm{Mpc}$. Here, $R_{200,\mathrm{AqA}}(z)$ is the virial radius of the Aq-A halo at redshift $z$ in comoving units. Haloes containing low-resolution dark matter particles are further excluded from our sample. There are $17$ and $10$ such haloes, which correspond to $0.7\%$ and $0.5\%$ of the total halo numbers, at $z=4.0$ and $2.5$ respectively. The total number of our selected haloes are summarized in the fourth column of Table \ref{table_gal_num}.

In Fig. \ref{fig_sim_visua}, we provide a visual impression of the formation of the Aq-A halo at $z=10.0,6.0,4.0,2.5,$ and $0.0$. From left to right column, we show dark matter overdensity, HI column density and gas temperature, respectively. The panels are centred on the Aq-A halo with a comoving scale $4 h^{-1}$Mpc. The circles around the centres of left panels indicate the virial radii of the halo at corresponding redshifts. At high redshifts, the proto Aq-A halo lies in intersection of a few massive filaments. This feature is quite common for massive galaxies at high redshifts \citep[e.g.][]{keres2005, dekel2006, ocvirk2008, dekel2009, keres2009, voort2011, danovich2012}. 

\begin{table} 
\centering
\caption{Number of haloes in the halo sample at different redshifts}\label{table_gal_num}
  \begin{tabular} {@{}cccc@{}}
  \hline
  $z$ & Filament & Field & Total\\
 \hline
   $4.0$ & $292$ & $2206$ & $2498$\\
   $2.5$ & $253$ & $1996$ & $2249$\\   
\hline
\end{tabular}
\end{table}

\subsection{Filament identification} \label{subsec_fila_iden}

\begin{figure*} 
\centering\includegraphics[width=400pt]{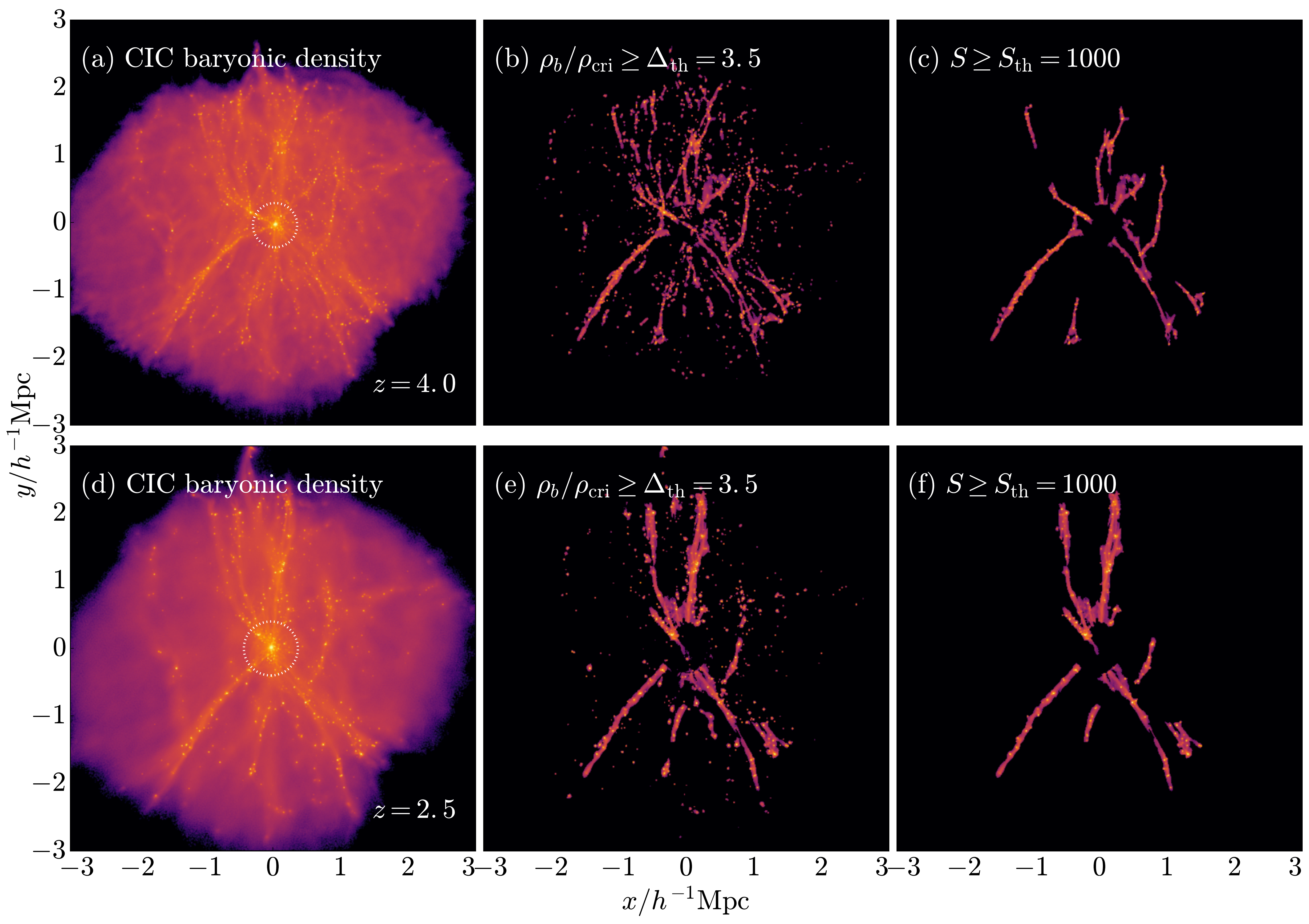} 
\caption{Filament identification at $z=4.0$ and $2.5$. Panels (a-c) illustrate the $xy-$projected baryonic density field of the zoom-in region at $z=4.0$ computed with the CIC assignment method, the grid cells with $\rho_b/\rho_\mathrm{cri} \geq \Delta_\mathrm{th}=3.5$, and the groups linked by the Hoshen-Kopelman algorithm with sizes $S \geq S_\mathrm{th}=1000$ respectively. The dotted circle in Panel (a) marks the spherical region with radius $r=2R_\mathrm{200,AqA}$ centred at the Aq-A halo. This spherical region is excluded when identifying filaments in Panels (b) and (c). Panels (d-f) show similar plots at $z=2.5$.}\label{fig_fila_iden}
\end{figure*}

This study is focused on the galaxy formation in filaments, thus it is essential to identify filaments from the simulation. As we are most interested in investigating the baryonic physics in filaments, our identification procedure is based on the baryonic density and is described as follows.

(i) Assign the baryonic (gas and star particles) density onto a $512^3$ dimensional grid with the Cloud-In-Cell (CIC) method. An illustration of the baryonic density fields at $z=4.0$ and $2.5$ is presented in Panels (a) and (d) of Fig. \ref{fig_fila_iden} respectively.

(ii) Exclude the spherical region with radius $r=2R_\mathrm{200, \mathrm{AqA}}$ from the Aq-A halo centre as the main halo is not of interest in this study. The excluded spheres at $z=4.0$ and $z=2.5$ are marked with dotted circles in Panels (a) and (d) of Fig. \ref{fig_fila_iden} respectively.    

(iii) Select the grid cells with densities satisfying $\rho_b(\bmath{r}_i)/\rho_\mathrm{cri} \geq \Delta_\mathrm{th}$, where $\Delta_\mathrm{th}$ is a free parameter to represent the baryonic overdensity. We adopt a value of $3.5$ for $\Delta_\mathrm{th}$, corresponding to a total matter overdensity of $\rho_m(\bmath{r}_i)/\rho_\mathrm{cri} \approx 20$. A projection on the $xy$ plane of the selected grid cells can be found in Panels (b) and (e) of Fig. \ref{fig_fila_iden}, where we can see some continuous filamentary structures and many small and isolated patches. These small patches are those grid cells that contain isolated haloes.

(iv) Link the neighbouring grid cells selected in the procedure (iii) into groups with the Hoshen-Kopelman algorithm \citep[][]{hoshen1976}, which was originally put forward in percolation studies. For each group, we define its size, $S$, by counting how many grid cells it has. We only keep groups with sizes $S \geq S_\mathrm{th}$, and define them as filaments. The size threshold parameter $S_\mathrm{th}$ is set to be $1000$ in this study. A visualization of the identified filaments is shown in Panels (c) and (f) of Fig. \ref{fig_fila_iden}. If a grid cell belongs to a filament group, then it is defined as a \textit{filament} grid cell; otherwise it is a \textit{field} grid cell. If an object (e.g., particle, halo, etc.) has its centre in a filament/field grid cell, then it is called a \textit{filament}/\textit{field} object (e.g., particle, halo, etc.). The total number of haloes in filaments and fields at $z=4.0$ and $2.5$ are presented in the second and third column of Table \ref{table_gal_num}.

The identification method outlined above has two free parameters, $\Delta_\mathrm{th}$ and $S_\mathrm{th}$. The overdensity parameter, $\Delta_\mathrm{th}$, is similar to the virial overdensity parameter used in a spherical overdensity halo finder, which determines the boundary of an identified structure. The size parameter, $S_\mathrm{th}$, sets the minimum size for identified filaments and naturally excludes those isolated groups shown in Panels (b) and (e) of Fig. \ref{fig_fila_iden}. These two parameters have clear physical meanings, and their values are set empirically. In this study, we choose $(\Delta_\mathrm{th},S_\mathrm{th})=(3.5, 1000)$, which generates clear filamentary structures according to human eyes' judgements. We have performed parallel analysis with other parameter sets (e.g., $\Delta_\mathrm{th}=2.5$ or $5.0$, and $S_\mathrm{th}=500$), and confirmed that our results are not sensitive to the chosen values of $\Delta_\mathrm{th}$ and $S_\mathrm{th}$.

\section{Results}\label{sec_res}

\subsection{Gas accretion in the residing low-mass haloes in filaments} \label{subsec_illust}
\begin{figure*} 
\centering\includegraphics[width=430pt]{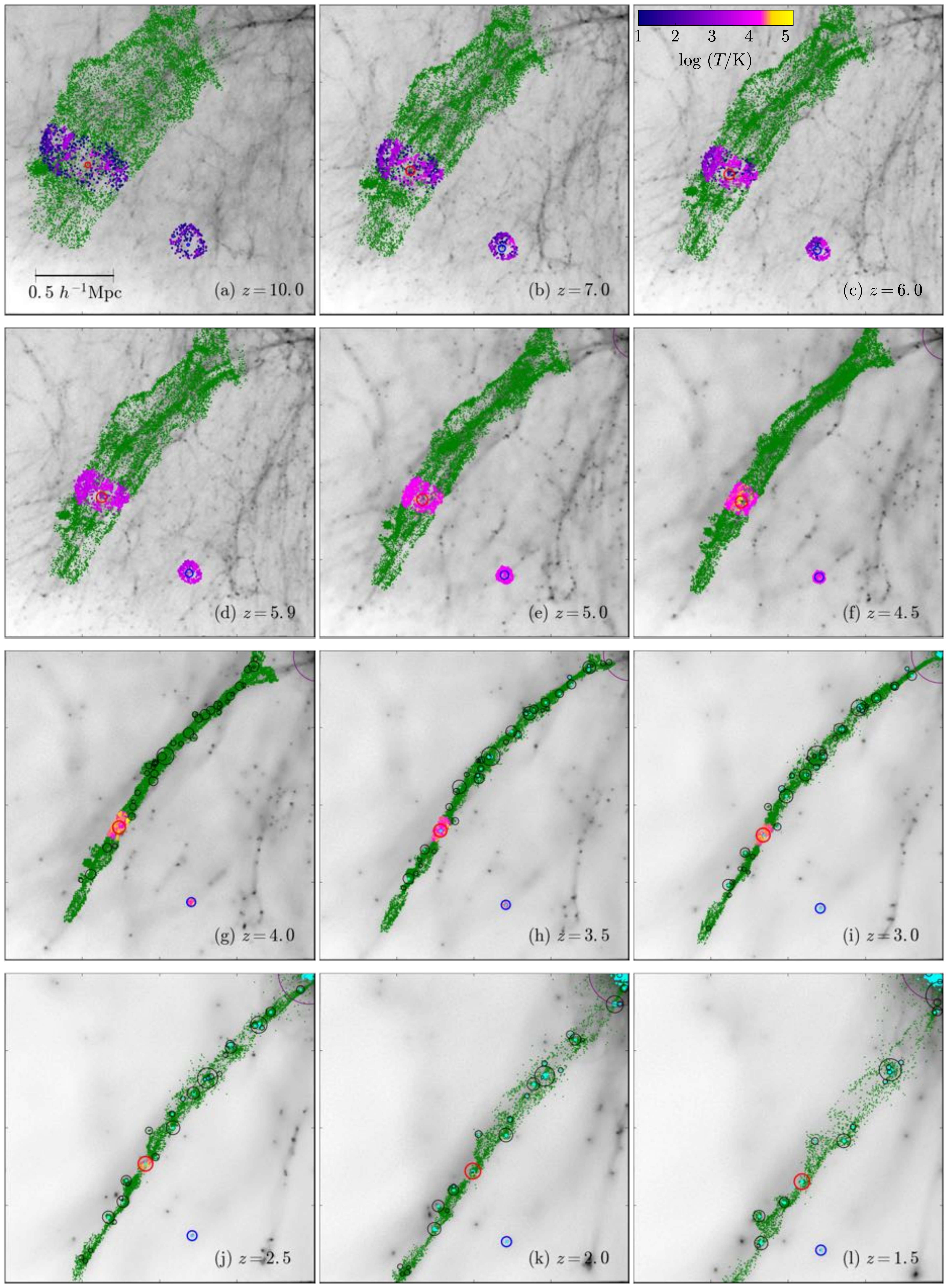} 
\caption{Illustration of the process that gas particles are accreted into a filament, and subsequently into dark matter haloes, and finally they form stars. We select a filament in quadrant III of the $xy$-projected baryonic density at $z=4.0$ shown in Fig. \ref{fig_fila_iden}. To see how the gas is accreted into this filament and its fate, we select those gas particles that are newly accreted into the filament at $z=4.0$ and mark them with green dots (Panel (g)). We then trace the selected particles back to higher redshifts (Panels (a-f)), and follow their later evolution (Panels (h-l)). The black circles mark the filaments haloes ($z \leq 4.0$) with masses $M_{200} > 10^8 h^{-1}M_\odot$, and the radius of each circle is set to the halo virial radius. The purple circles mark the Aq-A halo. The redshifts are labelled at the lower right corner of each panel. To see how these gas particles are accreted into haloes in the filament, we select a dark matter halo within the filament (the halo {\#}25, red circle), and mark the gas particles that will be accreted into it during the evolution with colors hue changes as temperature. The color bar is shown in Panel (c). As a comparison, we also select a field halo (the halo {\#}114, blue circle), and trace the evolution of the gas particles whose local overdensities just reach $\Delta_\mathrm{th}=3.5$ at $z=4.0$. Once a gas particle turns into a star, it is marked as cyan color. The comoving size of the $xy-$projected cubic region in each panel is $2$ $h^{-1}\mathrm{Mpc}$. The gray maps show the dark matter density.}\label{fig_gas_acc_fila}
\end{figure*}

To illustrate the gas accretion in filaments, in Fig. \ref{fig_gas_acc_fila}, we select the newly accreted particles from a pronounced filament at quadrant III of the projected $xy$ plane at $z=4.0$, and project their positions at several recorded redshifts. These ``newly accreted'' particles are identified as members of the filament at $z=4.0$ but not at the last adjacent recorded snapshot. These particles are shown as green dots on the top of a gray-scale dark matter density map in panel (g).  We also trace the motions of these gas particles before and after $z=4.0$ to illustrate their evolution. To see whether these gas particles can later join into individual haloes in the filament, we plot the filament haloes with masses more massive than $10^8 h^{-1}M_\odot$ with black circles with sizes scaled to their $R_{200}$. Cleary, these particles were quite diffusely distributed around the filament at early epochs and later were perpendicularly accreted into the filament. After these particles joined the filament, they rapidly condensed to the filament centre. Some of them were further accreted into filament haloes and eventually turned into stars (cyan dots) at lower redshifts, whereas some penetrated into the Aq-A halo centre through ``cold accretion''. 

To illustrate how the newly accreted filament gas particles can later join in filament haloes, we choose a filament halo {\#}25 at $z=2.5$ (marked as a red circle in panel (j)), and identify its gas particles which were newly associated with the filament at $z=4.0$, and trace them forward and backward to see their evolution. The positions of these particles are plotted as colored dots at $z=4.0$, and their color hues change as time according to their temperatures. One can clearly see that the gas accretion of the halo is quite anisotropic once they join into the filament, with the direction along the filament.  At $z=1.5$, almost all of these gas particles acquired at $z=4.0$ have turned into stars, which relates to the fact that we do not include feedback in the simulation. For a comparison, in the same figure, we show an example of the formation of a nearby field halo {\#114} (marked with a blue circle in Panel (g)). The halo has a virial mass of $2 \times 10^9$ $h^{-1}M_\odot$, which is similar to that of the halo {\#25}.  To have a fair comparison with the filament halo, we highlight the gas particles associated with the halo {\#}114 with exactly the same criterion according to local overdensity, namely requiring their densities below $\Delta_\mathrm{th}=3.5$ at $z=4.1$ but exceed this threshold at $z=4.0$. Apparently, the gas accretion of the field halo is quite isotropic.

\begin{figure*} 
\centering\includegraphics[width=500pt]{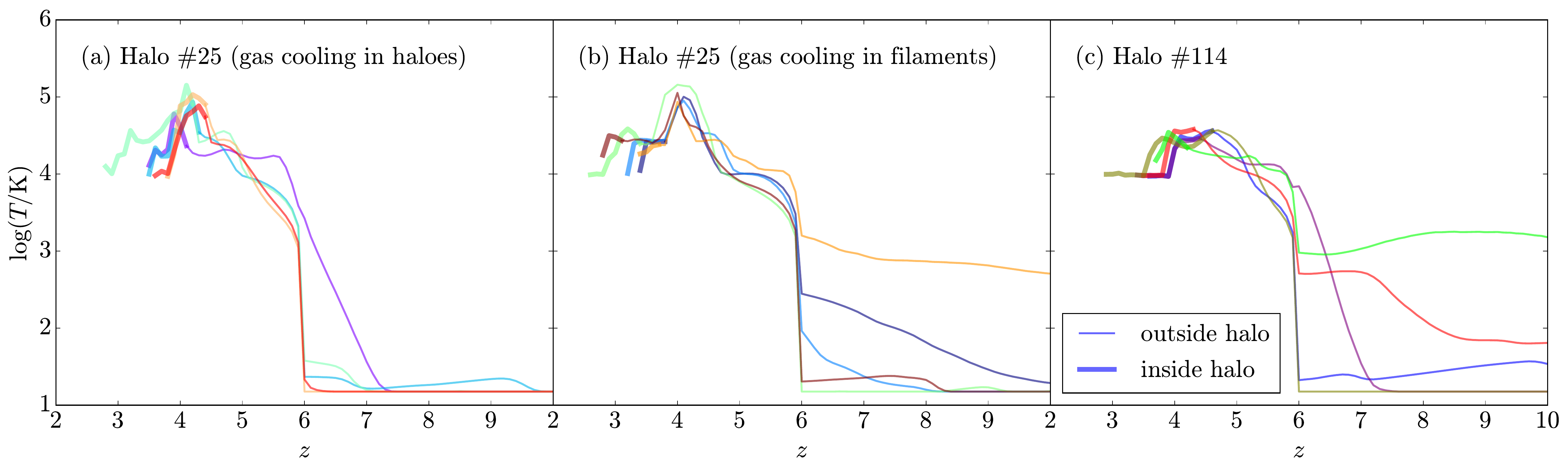} 
\caption{Temperature evolution of representative gas particles in the filament halo {\#}25 and field halo {\#}114.  There are two cooling modes for the gas particles in the halo {\#}25, which are displayed in Panels (a) and (b). Panel (a) shows five gas particles which cool down inside the halo. Panel (b) shows five gas particles which cool down before being accreted into the halo. In Panel (c), we plot 5 gas particles from the halo {\#}114. The thin (thick) line marks the gas particles outside (inside) the halo during their evolution. Different colors distinguish different gas particles. We plot their temperature histories from $z=10.0$ to the redshifts when they turn into stars. }\label{fig_temp_history}
\end{figure*}

From the above images, comparing with the field halo, gas accretion is more anisotropic for the filament halo (see Appendix \ref{ap_shock} for a further quantitative comparison between filament and field haloes). Apart from this, in addition to the classic gas cooling for a halo, the filament halo also acquires part of its cold and dense gas preprocessed by its surrounding filament environment. In Fig. \ref{fig_temp_history}, we illustrate clearly two different gas cooling modes in filament haloes. For the halo {\#25}, we first divide the newly accreted gas particles into two subsamples by considering whether a gas particle reaches its maximum temperature, $T_\mathrm{max}$, inside the halo. Then we select 5 random gas particles from each subsample, and plot their temperatures as a function of redshift in Panels (a) and (b) of Fig. \ref{fig_temp_history}, respectively. Note that we use thin (thick) lines to distinguish gas particles which lie outside (inside) $R_{200}$ of the halo. In Panel (a), all of these 5 particles reach their maximum temperatures inside the halo {\#}25 and cool down gradually to form stars. But for the gas particles in Panel (b), they are shocked by the filament at $z\sim 4$ (see Fig. \ref{fig_temp_map_halo} in Appendix \ref{ap_shock} for a visualization of the filament shock front), and thus reach their maximum temperatures outside the halo. Before they are accreted into the halo {\#}25, they have cooled down to $\sim T_\mathrm{max}/3$ in the filament. As a comparison, we also randomly select 5 newly accreted gas particles from the field halo {\#114}, in which the gas cooling is expected to be a singel mode. The thermal histories of these 5 particles are shown in Panel (c) of Fig. \ref{fig_temp_history}. As we expected, they reach their maximum temperatures inside the halo, and cool down later until they form stars, in line with the classic picture of gas accretion into a halo.

Note that the above discussions have mostly focused on two representative haloes, the filament halo {\#}25 and field halo {\#}114, but similar gas accretion processes can also be observed in the rest of filament and field samples.

\subsection{Fraction of gas preprocessed by filaments} \label{subsec_quanitify}
As discussed in the last subsection, for filament haloes, some of their gas particles are preprocessed by the filament before joining into them. An immediate question is, for a filament halo, what is the fraction of gas experiencing such filament preprocessing. As demonstrated in Fig. \ref{fig_temp_history}, the gas experiencing filament preprocessing has its $z_\mathrm{max}$ (i.e., the redshift that the gas particle reaches its maximum temperature) earlier than its $z_\mathrm{acc}$ (i.e., the redshift that the gas particle is accreted into a halo). Therefore, we can use the relation between $z_\mathrm{acc}$ and $z_\mathrm{max}$ to roughly classify gas particles into two categories. If $z_\mathrm{acc} \geq z_\mathrm{max}$, the gas is accreted into the halo without filament preprocessing; while if $z_\mathrm{acc} < z_\mathrm{max}$, the gas is firstly preprocessed in filaments and later accreted into the halo.

\begin{figure} 
\centering\includegraphics[width=240pt]{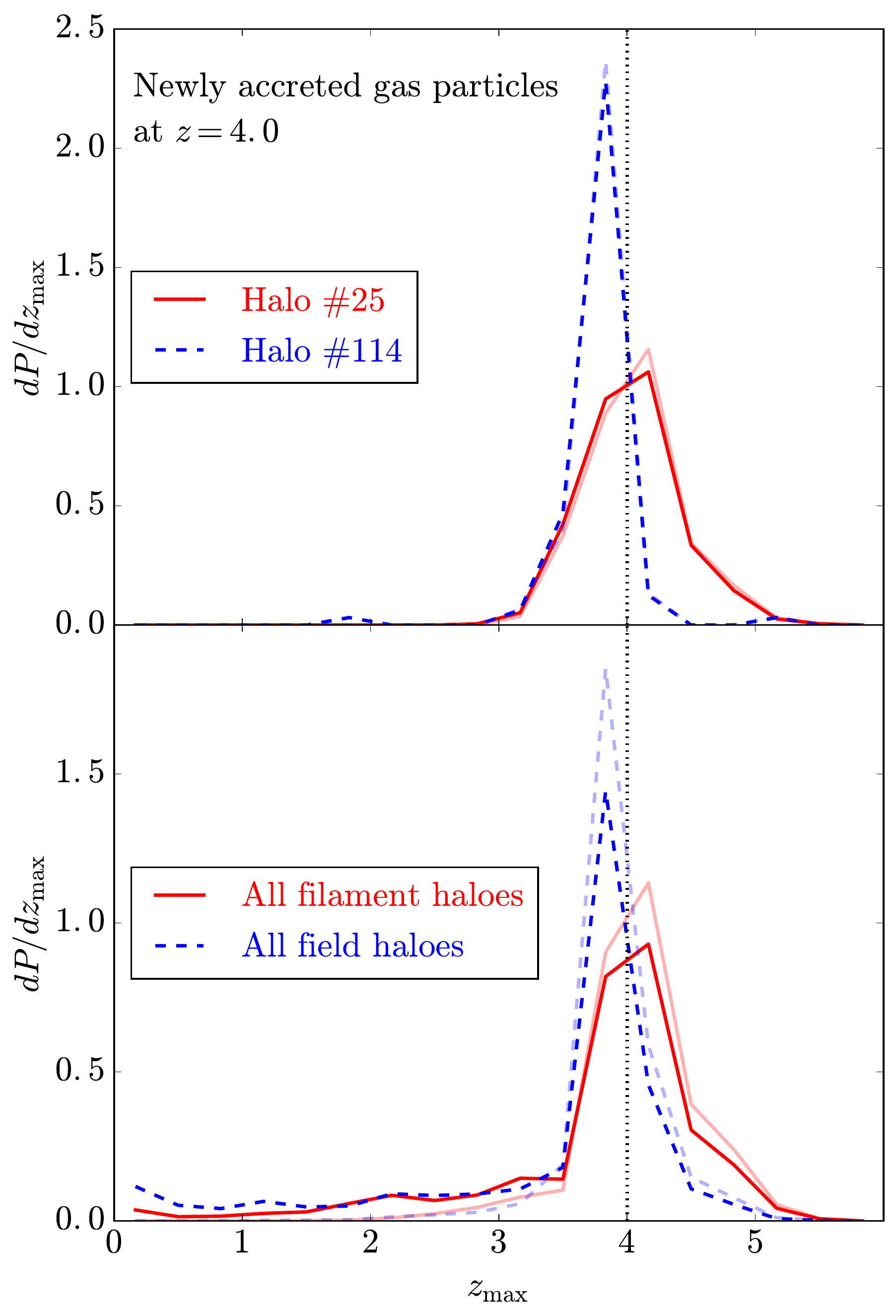} 
\caption{Top: Probability distribution functions of $z_\mathrm{max}$ for newly accreted gas at $z=4.0$ of the filament halo {\#}25 (red solid) and field halo {\#}114 (blue dashed). The lines with light colors show the case after excluding gas particles that experience several times of passing through and being re-accreted into haloes before or after $z=4.0$. The black dotted line marks the gas accretion redshift $z_\mathrm{acc}=4.0$. Bottom: similar to the top panel, but for all newly accreted gas at $z=4.0$ in all filament haloes (red solid) and field haloes (blue dashed).}\label{fig_zmax_two_gal}
\end{figure}

In the top panel of Fig. \ref{fig_zmax_two_gal}, we present the probability distribution functions (PDFs) of $z_\mathrm{max}$ for the newly accreted gas particles at $z=4.0$ (i.e. gas particles which are accreted into the halo between $z=4.1$ and $z=4.0$) for the filament halo {\#25} and field halo {\#114}. Clearly, the PDFs of the two haloes are quite different. For the field halo {\#114}, most of its gas particles reach their $T_\mathrm{max}$ after $z=4.0$ as expected in the classic picture of gas accretion into a halo. While, for the filament halo {\#25}, almost half of its gas particles reach $T_\mathrm{max}$ before accretion, and the rest half follows the standard picture. The results are similar if we plot the distributions for all newly accreted gas particles at $z=4.0$ in all filament and filed haloes; see the bottom panel of Fig. \ref{fig_zmax_two_gal}.

The fraction of gas preprocessed by filaments can be computed as
\begin{equation}
f_\mathrm{out} \equiv \frac{\Delta M_\mathrm{gas, out}}{\Delta M_\mathrm{gas}},
\end{equation}
where $\Delta M_\mathrm{gas}$ is the total mass of newly accreted gas, and $\Delta M_\mathrm{gas, out}$ is the mass of the newly accreted gas with $z_\mathrm{max}>z_\mathrm{acc}$ (i.e., the gas that reaches its $T_\mathrm{max}$ \textit{outside} the halo). For the newly accreted gas in the halo {\#}25 and halo {\#}114 at $z=4.0$, $f_\mathrm{out}$ are $44.7\%$ and $3.1\%$, respectively.

Note that some gas particles may pass through and are later re-accreted to the host halo more than once, and thus have complicated thermal histories. While it does not affect the results significantly by including or excluding them (see the curves with light colors in Fig. \ref{fig_zmax_two_gal}), we remove them from our sample.

\begin{figure*} 
\centering\includegraphics[width=500pt]{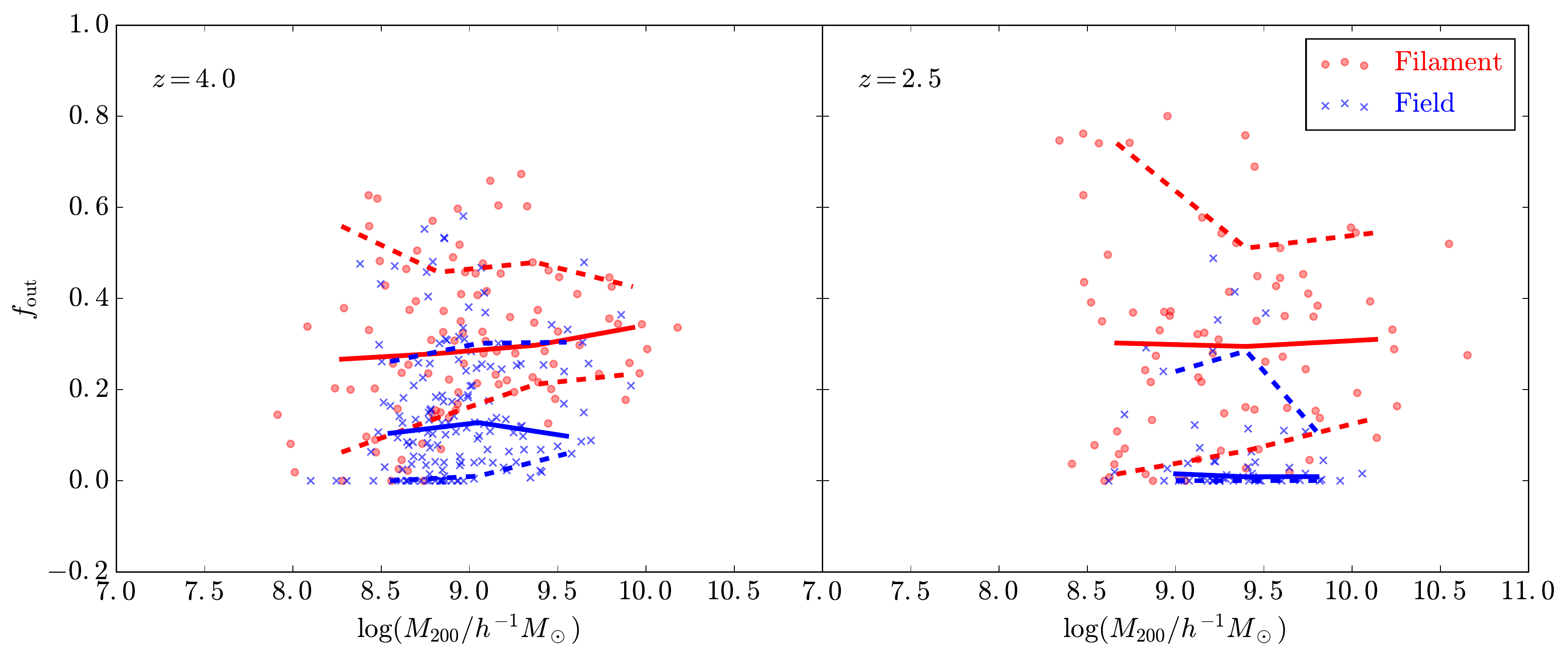} 
\caption{The fraction of gas preprocessed by filaments as a function of halo mass at $z=4.0$ (left) and $z=2.5$ (right). The filament and field haloes are plotted with red dots and blue crosses, respectively. The median of $f_\mathrm{out}$ in each mass bin are shown with solid lines. Dashed lines show the $15.9\%$ and $84.2\%$  percentiles of each mass bin.}\label{fig_fout_mass}
\end{figure*}

Similarly, we can calculate $f_\mathrm{out}$ for all filament haloes in our sample. In order to have a robust estimation, we only consider those haloes that have at least $100$ newly accreted gas particles. Also, in order to achieve  better statistics, we here define the newly accreted gas particles at $z_\mathrm{acc}$ as those that belong to haloes at $z_\mathrm{acc}$ but are not associated to any haloes at $z_\mathrm{acc}+0.3$. In Fig. \ref{fig_fout_mass}, we plot $f_\mathrm{out}$ as a function of $M_{200}$ for our halo sample at $z=4.0$ and $2.5$. When calculating the median and scatter of $f_\mathrm{out}$ in each mass bin, we require each mass bin to contain at least $10$ haloes. Overall, filament haloes (red dots) tend to have much larger  $f_\mathrm{out}$ than field haloes (blue dots). The median values of $f_\mathrm{out}$ for both filament and field haloes seem to be independent of halo mass. At $z=4.0$, the median value of $f_\mathrm{out}$ for filament haloes is about $30$ percent, while it is about $10$ percent for field haloes. Results for $z=2.5$ are similar. Note, at $z=4.0$, some field haloes have quite large $f_\mathrm{out}$. Our examination shows that these  haloes usually reside in smaller filaments with sizes $S<S_\mathrm{th}$.

\subsection{Impacts of filaments on halo properties}\label{subsec_impact}
\begin{figure*} 
\centering\includegraphics[width=440pt]{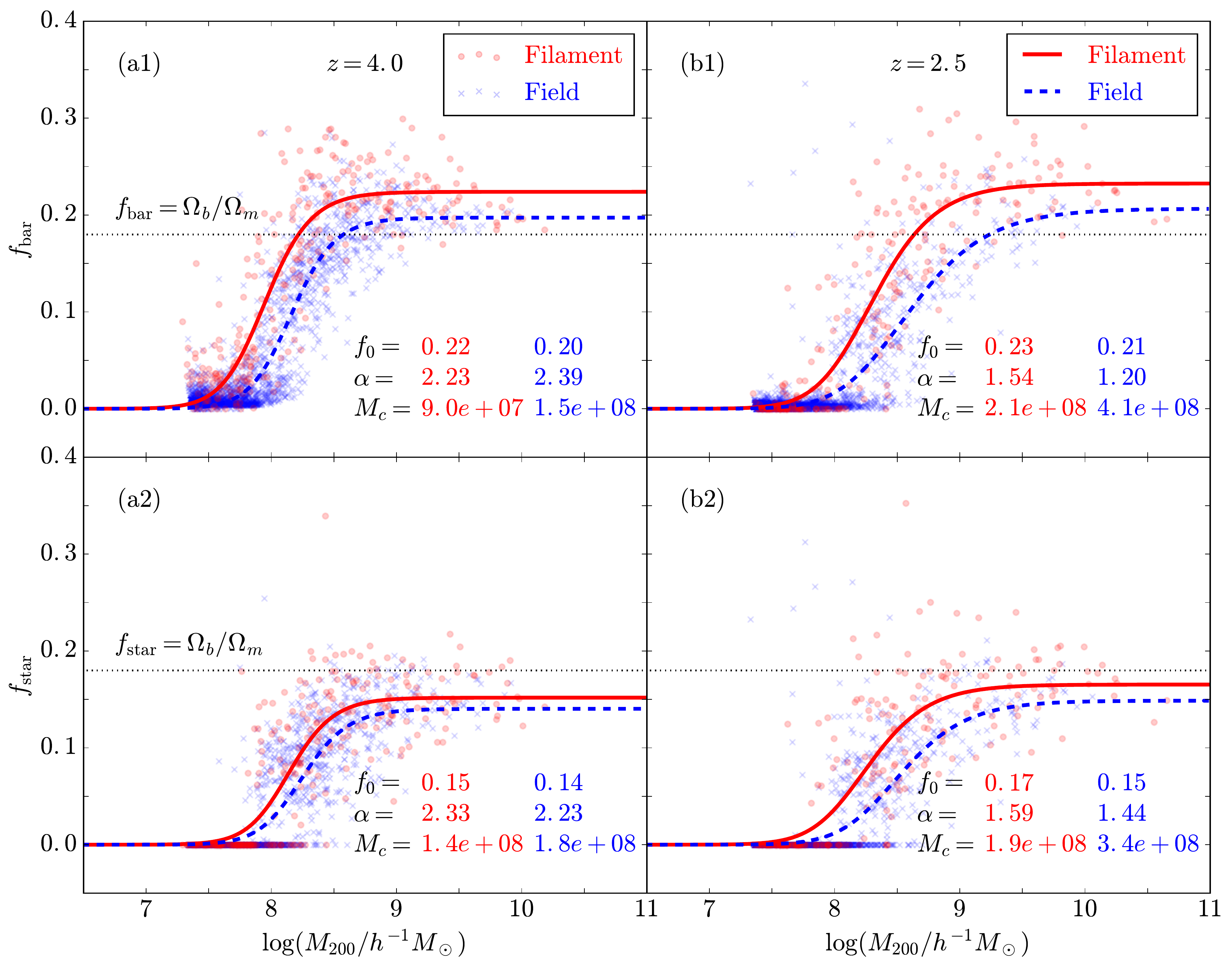} 
\caption{Baryon fractions (upper) and star fractions (lower) of haloes in filaments (red) and fields (blue) at $z=4.0$ (left) and $z=2.5$ (right). The light dots (crosses) plot the value of each filament (field) halo in our sample, and the solid (dashed) line shows the best-fitted curve according to Eq. (\ref{eqn_gnedin}). The fitted parameters are labelled at the lower right corner of each panel. The physical unit of $M_c$ is $h^{-1}M_\odot$.}\label{fig_gal_prop}
\end{figure*}

As shown in previous subsections, filaments significantly affect the intergalactic medium and gas accretion of the residing dark matter haloes. In this subsection, we provide comparisons of some general properties between field and filament haloes.  In Fig. \ref{fig_gal_prop}, we plot  baryon fractions $(f_\mathrm{bar})$ and star fractions $(f_\mathrm{star})$ as a function of virial masses for the filament and field haloes at $z=4.0$ and $2.5$. Results for  filament and field haloes are shown with dotted and crossed points, respectively. Here, the baryon fraction of a halo is defined as the ratio between the baryonic and virial mass, i.e., $f_\mathrm{bar}\equiv M_\mathrm{bar}/M_{200}$, and the star fraction is the ratio between the stellar mass and halo mass (i.e., $f_\mathrm{star}\equiv M_\star/M_{200}$).

To fit the relations between $f_\mathrm{bar}$ (or $f_\mathrm{star}$) and $M_{200}$, we adopt the function proposed by \citet[][]{gnedin2000} with an additional free parameter $f_0$,
\begin{equation}\label{eqn_gnedin}
f_\mathrm{bar,star}(M_{200}) = f_0\left[1 + \left(2^{\alpha/3} - 1\right)\left(\frac{M_c}{M_{200}}\right)^{\alpha} \right]^{-3/\alpha}.
\end{equation}
There are three free parameters, i.e., $f_0$, $\alpha$ and $M_c$, in this fitting function. Here $f_0$ is the baryon/star fraction for high-mass haloes (i.e., $M_{200} \gg M_c$), $\alpha$ indicates how rapidly $f_\mathrm{bar,star}$ drops in the low-mass end, and $M_c$ marks the characteristic mass scale that $f_\mathrm{bar,star}$ drops to $f_0/2$. The best fits are shown with solid and dashed lines in each panel for filament and field haloes, respectively, and the fitting parameters are labelled as well. Note that here we have included all haloes with at least $100$ particles. We have parallelly performed the fits for haloes with at least $500$ particles, and confirmed that haloes with $\la 500$ particles (i.e. with lower resolutions) do not affect our conclusions presented in the following.

Upper two panels display the baryon fraction results. At $z=4.0$ ($z=2.5$), due to the photoionization by the UV background, haloes with $M_{200} <10^{8.5}$ $h^{-1}M_\odot$ ($M_{200} <10^{9}$ $h^{-1}M_\odot$) tend to have lower $f_\mathrm{bar}$ with decreasing $M_{200}$ \citep[see also e.g.][]{okamoto2008}. Interestingly, filament haloes tend to have higher $f_\mathrm{bar}$ than field haloes, and the difference is even larger at $z=2.5$.  Similar trends can also be found in \citet[][]{metuki2015}, while the differences are very small between the filament and sheet/void haloes at $z=0$. The much larger difference shown here is perhaps because of two facts. (i) The dark matter haloes studied here are at higher redshifts and have lower masses; (ii) we focus on the haloes residing in more pronounced filaments.

Bottom panels of the same figure show the star fraction as a function of halo mass at $z=4.0$ and $2.5$. At $z=4.0$, star fractions of filament haloes are slightly higher than those of field ones, and the excess is further enhanced at $z=2.5$. Especially, the enhancement of $f_\mathrm{star}$ in filament haloes is more significant for haloes with lower masses at $z=2.5$, implying that filament structures have a positive impact on star formation in low-mass haloes. \citet[][]{metuki2015} also found similar trends in their simulations, but again the difference is very small. In observations, it has been also found that when approaching the cylindrical center of a filament, galaxies at $z\lesssim 1$ tend to have higher stellar masses \citep[e.g.][]{alpaslan2016, chen2017, darvish2017, kuutma2017, malavasi2017, poudel2017, laigle2017}.

According to the best-fitted parameters labelled  in Fig. \ref{fig_gal_prop}, we can see that the filament halo sample has a smaller characteristic  mass $M_c$ than its field halo counterpart. Especially, for both baryon and star fractions at $z=2.5$, $M_c$ for filament haloes is about only half of that of field haloes, indicating that the lowest mass limit to form a galaxy is smaller for the filament halo than the field one.

\section{Conclusions}\label{sec_con}
In this study, we use a high-resolution zoom-in hydrodynamical simulation of the Aq-A halo to study the impact of filaments on galaxy formation in their residing low-mass haloes ($M_{200}<10^{11}$ $h^{-1}M_\odot$) at high redshifts (e.g. $z=4.0$ and $2.5$). To identify filaments from the simulation, we developed a method based on the Hoshen-Kopelman algorithm with two parameters, i.e., $\Delta_\mathrm{th}=3.5$ and $S_\mathrm{th}=1000$. This method is shown to work well in our simulation. Our main conclusions are summarized as follows.

Similar to dark matter haloes, dark matter filaments can also trap and compress gas. As a result, at high redshifts, the accreted gas is shock heated at the boundary of filaments, cools rapidly and condenses into filaments centre. Some of the preprocessed cold and dense gas is further accreted into residing individual low-mass haloes in a direction along the filaments. Consequently, comparing with field haloes, gas accretion is very anisotropic for filament haloes, and there are two cooling modes for the gas of filament haloes. About 30 percent of the accreted gas of a residing filament halo was preprocessed by filaments, and this fraction is independent of halo mass. Filament haloes have higher baryon fractions and enhanced stellar fractions when comparing with their field counterparts. Our results thus suggest that filaments can assist gas cooling and enhance star formation in their residing dark matter haloes.  

Note, as a first step, this study is focused on investigating gas accretion and cooling in filaments and their residing low-mass dark matter haloes, and our simulation neglects stellar feedback. We expect that our conclusions may be quantitatively changed when including stellar feedback in the simulation, as some previous simulated works showed that feedback can affect halo gas accretion and star formation \citep[e.g.][]{oppenheimer2010, faucher_giguere2011, voort2011, murante2012, woods2014, nelson2015}. We will explore the effects of feedback on our results in our future work. 

\section*{Acknowledgements}
We thank the anonymous referee for the valuable comments. We thank Jie Wang, Shi Shao and Marius Cautun for discussions. LG acknowledges support from the NSFC grant (Nos 11133003, 11425312) and a Newton Advanced Fellowship, as well as the hospitality of the Institute for Computational Cosmology at Durham University. The simulations used in this work were performed on the ``Era" supercomputer of the Supercomputing Centre of Chinese Academy of Sciences, Beijing, China.

\appendix
\section{Shock fronts and gas accretion temperatures of Haloes {\#}25 and {\#}114} \label{ap_shock}

\begin{figure*} 
\centering\includegraphics[width=375pt]{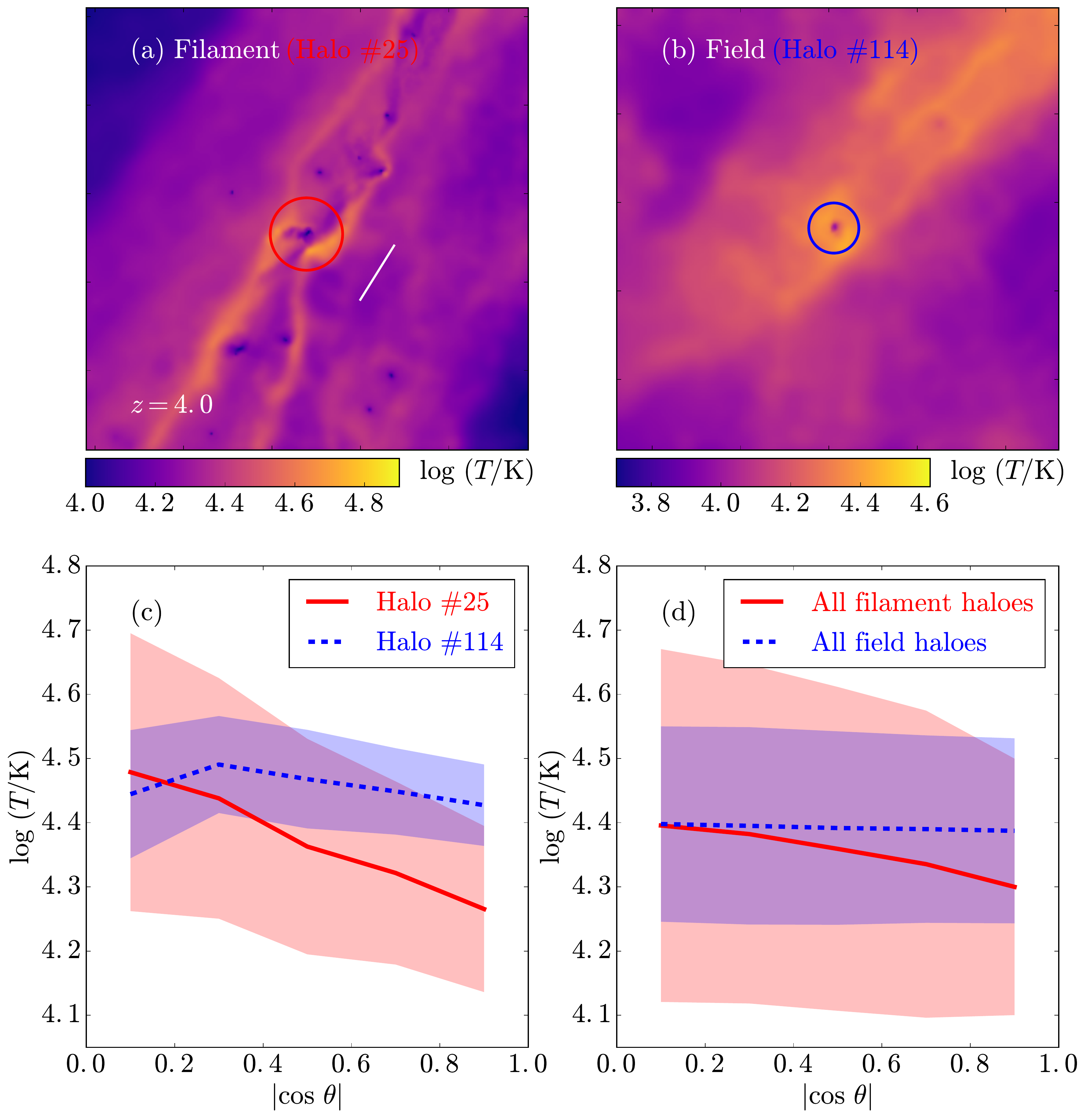} 
\caption{Gas temperature of a filament halo (the halo {\#}25, red) and a field halo (the halo {\#}114, blue) at $z=4.0$. In Panels (a) and (b), we plot the $xy-$projected cubic region with comoving length of $0.5$ ${h^{-1}\mathrm{Mpc}}$ centering at the halo {\#}25 and {\#}114 respectively. The circles mark the halo virial radii. The color bars are shown at the bottom of Panels (a) and (b). The white line segment indicates the filament direction at the center of the halo {\#}25; see the text for details. In Panel (c), we use the solid (dashed) line to plot the median temperature of the newly accreted gas particles in the halo {\#}25 ({\#}114) as a function of the accretion angle. The shaded region represents the $1\sigma$ scatters. Panel (d) is similar to Panel (c), but for all filament (solid) and field (dashed) haloes. See the text for details.}\label{fig_temp_map_halo}
\end{figure*}

In this appendix, we plot the gas shock fronts associated with the haloes {\#}25 and {\#}114, and quantitatively compare the different accretion behaviours of these two selected haloes. The results are shown in Fig. \ref{fig_temp_map_halo}.

In Panel (a), we plot the $xy$-projected temperature field of a cubic region with length of $0.5$ $h^{-1}\mathrm{Mpc}$ centering at the halo {\#}25 at $z=4.0$. As we can see from the figure, there is clearly a shock front at the surface of the filament. Behind the shock front, the gas temperature gradually increases when approaching the filament. Inside the filament, the gas temperature is lower than the shock front, which is due to gas cooling by filaments. Similar descriptions can be found in \citet[][]{gao2015}. At the center of the halo {\#}25, where is the star forming region, the gas temperature is as low as $\sim 10^4$ $\mathrm{K}$. Quantitatively, we can look at the relation between the temperature and the accretion direction of the accreted gas. To do so, we select the newly accreted gas particles for the halo {\#}25, i.e., those gas particles that are inside the halo {\#}25 at $z=4.0$ but they are not belong to the progenitor of the halo {\#}25 at $z=4.1$, and calculate their accretion angles, $\theta$, with respect to the filament direction. The accretion angle is defined as the angle between the vector from the halo center to the gas particle, $\bmath{r}_{i}^\prime =\bmath{r}_i-\bmath{r}_c$, and the filament direction $\bmath{r}_f$, i.e., 
\begin{equation}
\theta \equiv \arccos \left(\frac{\bmath{r}_{i}^\prime \cdot \bmath{r}_f}{|\bmath{r}_{i}^\prime||\bmath{r}_f|} \right).
\end{equation}
Here, $\bmath{r}_c$ and $\bmath{r}_i$ are the position vectors of the halo center and the $i$-th gas particle respectively, and the filament direction $\bmath{r}_f$ is calculated by solving the eigensystem of the Hessian matrix \citep[e.g.][]{hahn2007, zhang2009},
\begin{equation}
H_{ij}=\frac{\partial^2 \rho_s(\bmath{r})}{\partial x_i \partial x_j},
\end{equation}    
where $\rho_s(\bmath{r})$ is the smoothed filament density field with a smoothing scale of $0.25$ $h^{-1}\mathrm{Mpc}$. We have tested that our results are not sensitive to the value of the smoothing scale. The vector $\bmath{r}_f$ is the eigenvector that associates with the positive eigenvalue. The filament direction at the center of the halo {\#}25 is plotted as a white line segment in Panel (a). The median relation between the temperature and the accretion angle for the newly accreted gas particles of the halo {\#}25 is shown as the red solid line in Panel (c), with the light red shaded region showing the $1\sigma$ scatters. We can see that in general, the temperature of the gas accreted perpendicularly to the filament is $\sim 1.7$ times of that for gas accreted along the filament. This is a clear evidence of the anisotropic accretion behaviour. 

As a comparison, we also plot a field halo, the halo {\#}114, in Fig. \ref{fig_temp_map_halo}. From Panel (b), we see that the temperature field is more isotropic around the halo {\#}114. We also find a low temperature region at the center of the halo {\#}114 where the gas is about to become stars. The correlation between the temperature and the accretion angle for the newly accreted gas particles of the halo {\#}114 is shown as the blue dashed line in Panel (c). Here, without loss of generality, we adopt the filament direction $\bmath{r}_f$ calculated above to compute the accretion angle $\theta$ for the gas in the halo {\#}114. As our expectations, the blue dashed line in Panel (c) is nearly flat, which implies an isotropic accretion. We also notice that the value of the blue line is close to that of the red line at $|\cos \theta| \approx 0$. This means that the gas accretion in field haloes is similar to that of filament haloes with similar masses in the direction perpendicular to the filament. Comparing to the filament halo, the temperature - accretion angle relation for the field halo has smaller scatters, which reflects the fact that the gas accretion in field haloes is more similar in different directions while the newly accreted gas particles in filament haloes have more complicated and diverse histories.

Although above we only look at two specific haloes, similar conclusions can be drawn for the newly accreted gas of all filament and field haloes, as shown in Panel (d). Here, again without loss of generality, we use the filament direction $\bmath{r}_f$ of halo {\#}25 to compute the accretion angle $\theta$ for the gas in field haloes. We have also tested with the following used directions: (i) other global directions, e.g. $\bmath{r}=(1,0,0)$, (ii) eigenvectors computed from the local smoothed Hessian matrix for each field halo, which reflect the local matter flow directions at large scales, (iii) the filament direction of the nearest filament halo for each field halo, and all of these choices give similar results for field haloes (i.e. no anisotropic accretion is seen).

\label{lastpage}

\end{document}